\documentclass{cup-hpl}

\usepackage{color}
\usepackage[]{units}
\usepackage{amsmath}
\usepackage{slashed}

\newcommand{\eqnrefs}[2]{Eqs. (\ref{#1}) and (\ref{#2})}
\newcommand{\trm}[1]{\textrm{#1}}
\newcommand{\tsf}[1]{\textsf{#1}}
\newcommand{\be}{\begin{equation}}
\newcommand{\ee}{\end{equation}}
\newcommand{\bea}{\begin{eqnarray}}
\newcommand{\eea}{\end{eqnarray}}
\newcommand{\Ecr}{E_{\trm{cr}}}
\newcommand{\Icr}{I_{\trm{cr}}}
\newcommand{\Aext}{A_{\trm{ext}}}
\newcommand{\nvac}{\tsf{n}_{\trm{vac}}}
\newcommand{\mbf}[1]{\mathbf{#1}}

\newcommand{\eqnref}[1]{Eq.\ (\ref{#1})}
\newcommand{\eps}{\varepsilon}

\newcommand{\vphi}{\varphi}

\newcommand{\ket}[1]{| #1 \rangle}

\newcommand{\twovector}[2]{\left( \begin{array}{c}
                               #1 \\ #2 \end{array} \right)}

\newcommand{\figref}[1]{Fig.\ \ref{#1}}                               
\begin{document}

\newtheorem{theorem}{Theorem}

\shorttitle{Measuring Vacuum Interactions with High Power Lasers}                                  
 
\shortauthor{B. King}

\title{Measuring Vacuum Polarisation with High Power Lasers}

\author[1]{B. King \corresp{ \email{b.king@plymouth.ac.uk}}}
\author[1]{T. Heinzl}

\address[1]{Centre for Mathematical Sciences, Plymouth University, Plymouth, PL4 
8AA, United Kingdom}

\begin{abstract}
When exposed to intense electromagnetic fields, the quantum vacuum is expected to exhibit properties 
of a polarisable medium akin to a weakly nonlinear dielectric material. Various schemes have been 
proposed to measure such vacuum polarisation effects using a combination of high power lasers. 
Motivated by
several planned experiments, we provide an overview of experimental 
signatures that have been suggested to confirm this prediction of quantum electrodynamics of real photon-photon scattering. 
\end{abstract}


\maketitle

\section{Motivation}
The increasing availability of multi-hundred TW and PW lasers\cite{danson15} brings the 
confirmation 
of long-predicted phenomena of strong-field quantum 
electrodynamics\cite{marklund_review06,dipiazza12} (QED) closer. A multitude of effects on 
the polarisation, wave-vector and frequency of photons that probe the polarisation of 
the charged virtual pairs of the vacuum have been theoretically 
investigated. All of these 
effects can be understood in terms of the single process 
of ``photon-photon scattering''. The current best experimental limit on the 
predicted cross-section for photon-photon scattering using just high power laser pulses lies 
eighteen orders of magnitude above QED\cite{bernard00}, but recent laser-cavity experiments such as 
BMV\cite{rizzo13} and PVLAS\cite{DellaValle:2014xoa} have 
reduced this to six and three orders of magnitude respectively. Moreover, coinciding with the 
completion of the XFEL laser at DESY, an experiment at the HIBEF 
facility\cite{schlenvoigt15} plans to measure one manifestation of photon-photon scattering, namely 
the birefringence of the 
vacuum, using the XFEL beam and a $1\,\trm{PW}$ optical laser. This has generated much 
interest in 
vacuum polarisation effects. 

The aims of this work are two-fold. First, the main analytical approaches used to study 
photon-photon scattering will be shown to be essentially equivalent for predictions of planned 
laser experiments. Second, an overview of the predicted signatures of real 
photon-photon scattering in various experimental scenarios will be provided, which it is  
hoped will also be useful for the 
non-specialist and in particular promote discussions between theorists and experimentalists.
%
%
%
%
%
%

\section{Introduction: Vacuum Polarisation}

\begin{figure}[h]
\begin{center}
\includegraphics[width=0.4\linewidth]{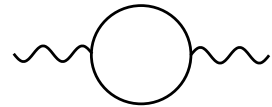}
\caption{\label{fig:VacPol}Vacuum polarisation loop in QED. Wavy and straight lines represent 
photons and fermions (electrons and positrons), respectively.}
\end{center}
\end{figure}

Vacuum polarisation, depicted in the Feynman diagram of Fig.\ \ref{fig:VacPol}, is a basic 
radiative correction that modifies the propagation of photons in vacuum through the appearance 
of virtual pairs 
in 
a `fermion loop'. 

There are two complementary interpretations of this effect. The first is based on what is called 
`old-fashioned' perturbation theory which emphasises energy considerations at the price of manifest 
covariance\cite{Weinberg:1995mt}. In this interpretation, Heisenberg's uncertainty relation is 
invoked to show how quantum mechanics predicts 
energy and momentum conservation may be violated. The amount of this violation is 
inversely 
proportional to the space-time scale over which it occurs. This effect is represented by 
short-lived 
``virtual'' particles. The second, equivalent interpretation is manifestly covariant and regards 
the 
virtual pairs as quantum fluctuations. In this interpretation, at any space-time point there is a 
non-vanishing probability 
amplitude for a photon to fluctuate into a pair (or a pair and a photon or in fact any number of 
particles allowed by the original photon quantum numbers). In this view, energy-momentum 
conservation 
is not violated, but the virtual particles do not obey Einstein's famous equation relating energy 
and mass. 

The main physical effect of vacuum polarisation is charge renormalisation due to polarisation 
screening as explained in any standard quantum field theory text\cite{peskin95}. The electric charge 
of a particle increases as one `dives' into its virtual polarisation 
cloud, hence with decreasing distance from the particle. As a result, the electric charge becomes 
scale-dependent which may be expressed in terms of a distance-dependent fine structure constant, 
$\alpha = \alpha(R)$. At distances large compared to the electron Compton wavelength, $R = 
\lambdabar_e = \hbar/mc$, the typical length scale of QED, one has $\alpha = e^2/4\pi\hbar c 
\simeq 
1/137$. However, at the much smaller Compton wavelength of, say, the $Z$ boson, $R = \lambdabar_Z = 
\hbar/M_Z c$, the QED coupling $\alpha$ increases to $\alpha(\lambdabar_Z) \simeq  1/128$.

\begin{figure}[h]
\begin{center}
\includegraphics[width=0.35\linewidth]{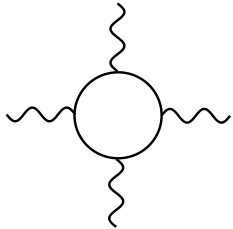}
\caption{\label{fig:PhotPhot}Probing vacuum polarisation by photon-photon scattering. }
\end{center}
\end{figure}

At typical laser energies, the dominant screening particles are indeed pairs of virtual electrons 
and positrons. Their (virtual) presence may be probed by coupling them to additional photons (see 
Fig.\ \ref{fig:PhotPhot}), which may represent either fluctuating quantum fields or classical 
background fields such as provided by lasers. In either case, we are led to consider the probing of 
vacuum polarisation by ``photon-photon scattering''. When large numbers of photons are involved, a 
classical metaphor of this quantum effect is of charged vacuum pairs forming a polarisable ``vacuum 
plasma'' medium with a nonlinear susceptibility and permeability. An important consequence of this 
quantum correction to Maxwell's equations is the violation of the principle of superposition for 
electromagnetic waves in vacuum.

%
%
%
%
%
%
\section{Analytical Methods}

The microscopic theory describing laser-matter or laser-laser interactions is QED described by the 
Lagrangian
\bea
  \mathcal{L}_{\trm{QED}} = \bar{\psi} (i \slashed{\partial} - m) \psi - 
  \frac{1}{4} F_{\mu\nu} F^{\mu\nu} - e \bar{\psi} \slashed{A} \psi \; ,
\eea
the separate terms representing the Dirac, Maxwell and interaction Lagrangians, respectively. The 
latter derives from `minimal substitution', that is the replacement of the ordinary by the covariant 
derivative, $i \partial \to i \partial - e A \equiv i D_A$ in the free Dirac term, which leads to 
the usual coupling of the photon field $A_\mu$ to the Dirac current $j^\mu = e \bar{\psi} \gamma^\mu 
\psi$ as $ e \bar{\psi} \slashed{A} \psi = A_\mu j^\mu$. An intense laser field will normally be 
included as a classical,  external background field $\Aext$ by the prescription of replacing 
$A \to A + \Aext$ in the interaction term only. This guarantees that $\Aext$ is not altered by
the interaction because the Maxwell term 
will only contain the field strength tensor built from the fluctuating fields $A_{\mu}$, i.e.\  
$F_{\mu\nu} = \partial_\mu A_\nu - \partial_\nu A_\mu$.

In this contribution we are interested in laser-laser interactions. In this case, the centre-of-mass 
energy (even for x-rays) will always be much lower than the electron rest energy, $mc^2$. It 
is thus 
sufficient to work with the low-energy effective field theory obtained from the QED Lagrangian by 
`integrating out' the Dirac fields.  This can be done by employing the functional integral 
representation of  the QED vacuum persistence amplitude $Z$ relating in and out vacua:
\bea
  Z &=& \int \mathcal{D}A\, \mathcal{D} \psi\, \mathcal{D} \bar{\psi} ~
  \exp \left(i S_{\trm{QED}}[A, \psi, \bar{\psi}]\right) \nonumber \\
  &\equiv&   \int \mathcal{D}A \, \exp \left(i S_{\trm{eff}} [A]\right) \; .
\eea
In the second step, the fermionic degrees of freedom have been integrated out by performing a 
Gaussian integral resulting in a fermionic determinant, 
\bea
  \exp \left(i S_{\trm{eff}} [A]\right) = 
  \exp \left(\trm{Tr} \ln \frac{i \slashed{D}_A - m}{i \slashed{\partial} - m} \right)\; ,
\eea
where we have re-exponentiated using $\trm{Det}  = \exp \trm{Tr} \ln$. The fermionic determinant 
depends on the photon field $A$ and can only be evaluated analytically for special configurations 
such as constant fields. Alternatively, one may perform a derivative (i.e.\ low-energy) 
expansion\cite{gusynin99, dunne99}, 
the leading order of which coincides with the constant field evaluation. For QED this has been done 
long ago (using different techniques) \cite{heisenberg36,weisskopf36,schwinger51} the result being 
the celebrated Heisenberg-Euler Lagrangian
\bea
  \mathcal{L}_{\trm{HE}} &=& 
  -\frac{m^{4}}{8\pi^{2}}\int_{0}^{\infty}\!\! ds\, 
  \frac{\exp\left(-s\right)}{s^{3}} 
  \Big[s^{2}ab\,\trm{cot}\,as\,\trm{coth}\,bs  \nonumber\\ 
  &&\qquad\qquad\qquad- 1+\frac{s^{2}}{3}(a^{2}-b^{2})\Big],\label{eqn:LEH}
\eea
where the dimensionless secular invariants $a$ and $b$ are given by:
\begin{align}
  a =
  \frac{\left[\sqrt{\mathcal{F}^{2}+\mathcal{G}^{2}}+\mathcal{F}\right]^{1/2}}{\Ecr};
  &\quad& b =
  \frac{\left[\sqrt{\mathcal{F}^{2}+\mathcal{G}^{2}}-\mathcal{F}\right]^{1/2}}{\Ecr}. \nonumber
\end{align}
These contain the two electromagnetic invariants 
\bea
  \mathcal{F} &=& -F_{\mu\nu}F^{\mu\nu}/4 = 
  (E^{2}-B^{2})/2 \; , \\
  \mathcal{G} &=& -F^{\mu\nu}\tilde{F}_{\mu\nu}/4 = 
  \mbf{E}\cdot\mbf{B} = 0 \; ,  
\eea
with field and dual field strength tensors, electric and magnetic fields ($F^{\mu\nu}$, 
$\tilde{F}^{\mu\nu}$, $\mbf{E}$ and $\mbf{B}$, 
respectively) and the critical field strength:
\bea
  \Ecr = m^{2}c^{3}/e\hbar \equiv \frac{m^2}{e} \; .
\eea
(Note that we now adopt natural units, $\hbar=c=1$, for the remainder of this section unless 
otherwise explicitly stated.) The critical, 
``Sauter'' \cite{Sauter:1931zz}  or ``Schwinger'' \cite{schwinger51} field-strength $\Ecr$ is built from 
the fundamental constants of QED and is the typical field-scale separating weak ($E\ll \Ecr$) from 
strong-field ($E > \Ecr$) vacuum polarisation phenomena.

\begin{figure}[h]
\begin{center}
\includegraphics[width=0.35\linewidth]{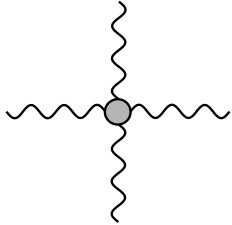}
\caption{\label{fig:PhotPhotHE}The leading order Heisenberg-Euler vertex or photon-photon scattering 
at low energies.}
\end{center}
\end{figure}

The Heisenberg-Euler Lagrangian (\ref{eqn:LEH}) is equivalent to QED for arbitrary values of the 
field strength but at energies small compared to $mc^2$. For the foreseeable future, laser 
experiments 
will stay well below the critical field strength, hence in the weak-field limit. Thus, to a very 
good approximation, it is sufficient to work with the leading order in a field strength expansion of 
(\ref{eqn:LEH}) given by:
\bea \label{LHE.LO}
  \mathcal{L}_{\trm{HE}}^{(2)} \simeq c_1 \mathcal{F}^2 + c_2 \mathcal{G}^2 \; ,
\eea
with dimensionless low-energy constants 
\bea
  \left\{ \begin{array}{c} c_1 \\ c_2  \end{array}  \right\} =  
  \frac{2 \alpha^2}{45 m^4} \times 
  \left\{ \begin{array}{c} 4 \\ 7  \end{array}  \right\} \; .
\eea
These define effective vertices corresponding to the low-energy limit of the diagram in Fig.\ 
\ref{fig:PhotPhot} with the fermion-loop no longer being resolved, see Fig.\ \ref{fig:PhotPhotHE}.

The cross section for the low-energy limit of real photon-photon scattering depicted 
in \figref{fig:PhotPhotHE} is given by\cite{landau4}:
\[
 \sigma = \frac{973}{10125\pi}\,\alpha^{4}\,
\left(\frac{\omega}{m}\right)^{6}~\lambdabar_{e}^{2};\qquad \omega \ll m
\]
whereas the high-energy limit is given by\cite{landau4}:
\[
 \sigma = 4.7\, \alpha^{4} \left(\frac{m}{\omega}\right)^{2}\,\lambdabar_{e}^{2}; \qquad \omega\gg 
m.
\]
The maximum of the cross-section is at the pair-creation threshold of colliding photon 
centre-of-mass energies $\omega = m$.

\subsection{Scattering Matrix}

\begin{figure}[h]
\begin{center}
\includegraphics[width=0.35\linewidth]{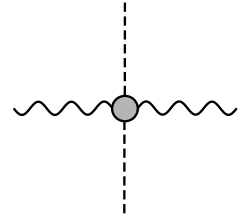}
\caption{\label{fig:PhotPhotExt}A probe photon (wavy lines) scattering off a classical laser 
background (dashed lines) at low energy (so that the Heisenberg-Euler vertex can be 
employed).}
\end{center}
\end{figure}

In what follows, we will consider a modification of the 4-photon scattering amplitude at low energy 
by assuming that two of the photons involved are stemming from a high intensity laser which is 
probed by a dynamical photon `passing through'. This is visualised in Fig.\ \ref{fig:PhotPhotExt}

We assume that an incoming probe photon with four-momentum $k$ and four-polarisation
$\varepsilon$ 
scatters off a laser background described by a field strength tensor $F_{\mu\nu}$ resulting in an 
outgoing photon with quantum numbers $k'$ and $\varepsilon'$. The resulting scattering amplitude is 
given 
by the $S$-matrix element
\bea
  \langle \varepsilon', k' ; \mbox{out} | \varepsilon, k ; \mbox{in}\rangle = 
  \langle \varepsilon', k'| \hat{S} | \varepsilon, k \rangle \equiv 
  S_{\trm{fi}} (\varepsilon', k', \varepsilon, k) \; .
\eea
Using the leading-order Lagrangian (\ref{LHE.LO}), writing $S_{fi}(\varepsilon', k', 
\varepsilon, k)$ as $S_{\trm{fi}}(q)$, the $S$-matrix element takes on the 
simple form of a Fourier integral
\bea
    S_{\trm{fi}}(q) = -i \! \int d^4 x \, e^{i q \cdot x} \,  S_{\trm{fi}}(x) \; ,  
\eea
where $q = k' - k$ is the momentum transfer and 
\be \label{eqn:SFI.X}
   S_{\trm{fi}}(x) =  c_1 (k', F\varepsilon')(k, F\varepsilon) + 
   c_2 (k', \tilde{F}\varepsilon')(k, \tilde{F}\varepsilon) \; ,
\ee
employing the abbreviated scalar products $(k, F\varepsilon) \equiv k_\mu F^{\mu\nu} \varepsilon_\nu$ etc. 
Hence,  one may introduce an \emph{intensity form factor},
\be \label{eqn:INTENSITY.FF}
  W^{\mu\alpha, \nu \beta} (q) \equiv -i \int d^4 x \, e^{i q \cdot x} 
  (c_1 F^{\alpha\mu} F^{\beta, \nu} + c_2 \tilde{F}^{\alpha\mu} 
  \tilde{F}^{\beta, \nu})\;, \\
\ee
which is the Fourier transformation of the background intensity distribution. 
In terms of the latter the scattering amplitude may be written as
\be \label{eqn:SFI.Q}
  S_{\trm{fi}}(q) = \varepsilon'_\alpha k'_\mu W^{\mu\alpha, \nu\beta} (q)
  k_\nu \varepsilon_\beta \; .
\ee
The results above are reminiscent of elastic electron nucleus scattering, where the scattering 
amplitude is proportional to the nuclear charge form factor which is nothing but the Fourier 
transform of the nuclear charge distribution. In photon-photon scattering one is naturally probing 
an intensity, rather than a charge, distribution. To proceed, one has to choose a suitable laser 
background field, $F_{\mu\nu}(x)$, and calculate its intensity form factor 
\eqnref{eqn:INTENSITY.FF}.

\subsection{Polarisation operator}

An equivalent representation is obtained in terms of a quantity aptly called the polarisation 
operator, denoted $\Pi^{\mu\nu}$. In its simplest incarnation it is just the mathematical expression 
for the Feynman diagram of \figref{fig:VacPol}, namely
\be
  \Pi^{\mu\nu} = -i e^2 \, \trm{tr}_\gamma \int \frac{d^4 p}{(2\pi)^4}
  \gamma^\mu \frac{1}{\slashed{p} - m}\gamma^\nu 
  \frac{1}{(\slashed{p} - \slashed{k} - m)} \, ,
\ee
where the trace $\trm{tr}_{\gamma}$ extends over the Dirac matrices $\gamma^\mu$. One may 
generalise this to the 
polarisation tensor in an external field $\Aext$, where one trades the free fermion propagators for 
interacting ones through the standard  minimal substitution $p \to p - e \Aext$. Indeed, this 
method has a long history\cite{Toll:1952rq,narozhny69,baier75a,meuren13} as reviewed by 
\cite{Dittrich:2000zu}. For our purposes it is sufficient to just employ the first-order 
weak-field Heisenberg-Euler Lagrangian (\ref{LHE.LO}) once again and rewrite it as 
\be 
  \mathcal{L}_{\trm{HE}}^{(2)} = \frac{1}{2} A_\mu \Pi^{\mu\nu} [\Aext]
   A_\nu  \; ,
\ee
with the polarisation tensor thus defining the second-order term.  From (\ref{LHE.LO}) one can 
straightforwardly read off that
\be \label{eqn:POL.TENSOR}
  \Pi^{\mu\nu} [\Aext] = \frac{c_1}{4} k_\alpha F^{\alpha\mu} F^{\beta\nu} k_\beta + 
  \frac{c_2}{4} k_\alpha \tilde{F}^{\alpha\mu} \tilde{F}^{\beta\nu} k_\beta \; ,
\ee
where the background field strength $F^{\mu\nu} = \partial^\mu \Aext^\nu - \partial^\nu \Aext^\mu$. 

To connect this approach with the $S$ matrix formalism we specialise to forward scattering by 
setting $k= k'$ in (\ref{eqn:SFI.X}) which yields the relation
\be \label{eqn:SFI.FWD}
  S_{\trm{fi, fwd}}(k)  = \varepsilon'_\mu(k) \Pi^{\mu\nu}(k) \varepsilon_\nu(k) \; . 
\ee
This makes the link between the polarisation operator and scattering matrix 
approaches manifest.

\subsection{Modified Maxwell Equations}

In standard quantum field theory notion\cite{peskin95}, the total Heisen\-berg-Euler action, 
$S_{\trm{eff}} = \int d^4 x \, \mathcal{L}_{\trm{eff}}$,  is nothing but the one-loop effective (or 
quantum) action of QED evaluated at low energies where there are no external electron lines. The 
associated effective Lagrangian is the sum of the classical Maxwell term 
$\mathcal{L}_{\trm{M}} = (E^{2}-B^{2})/2$ and the first quantum correction:
\bea
  \mathcal{L}_{\trm{eff}} = \mathcal{L}_{\trm{M}} + 
  \mathcal{L}_{\trm{HE}},
\eea
By variation of the quantum action, one can derive the corresponding modified Maxwell equations 
\cite{Akhiezer:1965}:
\bea
\pmb{\nabla}\cdot\mbf{E} = \rho_{\trm{vac}}; \quad \pmb{\nabla}\wedge \mbf{B} = 
\mbf{J}_{\trm{vac}} + \partial_{t} \mbf{E},
\eea
in which:
\bea
\rho_{\trm{vac}} = \pmb{\nabla}\cdot\mbf{P}_{\trm{vac}}; \quad  \mbf{J}_{\trm{vac}} = 
\pmb{\nabla}\wedge 
\mbf{M}_{\trm{vac}} + \partial_{t}\mbf{P}_{\trm{vac}}
\eea
and the vacuum polarisation and magnetisation are:
\bea
\mbf{P}_{\trm{vac}} = \frac{\partial\mathcal{L}_{\trm{HE}}}{\partial \mbf{E}}; \quad 
\mbf{M}_{\trm{vac}} = \frac{\partial\mathcal{L}_{\trm{HE}}}{\partial \mbf{B}}.
\eea
The wave equations:
\bea
\partial_{t}^{2}\mbf{E}-\pmb{\nabla}^{2}\mbf{E} &=& -\pmb{\nabla}\rho_{\trm{vac}}[\mbf{E},\mbf{B}] 
- \partial_{t}\mbf{J}_{\trm{vac}}[\mbf{E},\mbf{B}] \label{eqn:E1}\\
\partial_{t}^{2}\mbf{B}-\pmb{\nabla}^{2}\mbf{B} &=& 
\pmb{\nabla}\wedge\mbf{J}_{\trm{vac}}[\mbf{E},\mbf{B}],
\eea
can be solved using, for example, the method of Green's functions.
%
%
%
%
%
%
\section{Signatures of Vacuum Polarisation}
\begin{figure}[h!]
\centering
\includegraphics[width=0.4\linewidth]{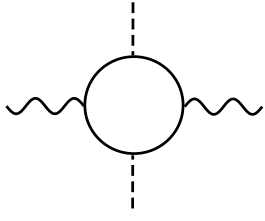}
\caption{Photons from the pump (dashes) interact with those from the probe to produce a 
pump-dependent vacuum index of refraction.} \label{fig:4ph1}
\end{figure}

The most general vacuum polarisation diagram represents an elastic scattering  amplitude that 
relates an incoming ensemble of photons $\ket{k_{1},\ldots,k_{n}}$, which interact in some 
experimental 
scenario, to an outgoing ensemble of photons $\ket{k'_{1},\ldots,k'_{n'}}$. In this review, we 
concentrate on processes that could be measured using high power lasers. The fields of these lasers 
are included in calculations in various ways. A ``monochromatic plane wave'' will refer to an 
infinitely extended wave with no transverse structure, a ``beam'' will refer to some 
inclusion of structure, e.g.\ a cylinder of radiation is a ``beam'', a 
``focussed beam'' will imply some approximation to a real beam with focal width as a parameter and a 
``pulse'' to a field localised in time with pulse duration as a parameter. 
Since laser pulse wavelengths are much larger than the Compton wavelength, and since 
expected electric field strengths are much less than the critical Sauter field, equivalent to an 
intensity of the order of $10^{29}\,\trm{Wcm}^{-2}$, the interaction of laser pulses with virtual 
electron-positron pairs can be expanded in terms of weak fields. Starting at $n=2$ as in 
\eqnref{LHE.LO}, each perturbative order describes a vacuum $2n$-wavemixing process. It is 
noteworthy that 
unlike when real electrons and positrons interact with intense laser fields, for virtual  
electron-positron pairs, the number of laser photons involved is typically small\cite{dipiazza13}, 
which is why the discussion is mostly in terms of four-wave mixing processes such as in
\figref{fig:4ph1}. This means the vacuum is often compared to a nonlinear optical material 
with a Kerr-like response\cite{delphenich06}. Although there is a large overlap with 
nonlinear optics, a major 
difference is that the polarisation of the dielectric (here, the vacuum), can be shaped by the 
pump laser pulse.

The majority of suggested signals of vacuum polarisation can be described by considering how the 
photons from a probe laser change due to interaction with a more intense pump laser. The pump laser 
will also be referred to as the ``background'' or the ``strong field'' where appropriate. The probe 
laser quantities will often be denoted with subscript $p$ and the pump or strong laser 
quantities with the subscript $s$. The source of probe photons will mostly be a high power 
laser, which, satisfying $E/\Ecr \gg \sqrt{\alpha}(\omega/m)^{2}$, often allows the external field 
concept to be invoked for the probe\cite{narozhny15}. Therefore the discussion will 
include interchangeably effects on probe photons and on the probe 
electromagnetic field, which assumes the photon-scattering process can be summed incoherently over 
the probe photon distribution. We begin by reviewing the consequence of real photon-photon 
scattering at the level of probe laser photons:
\bea
\gamma(\omega, \mbf{k},\eps(k)) \to \gamma(\omega', \mbf{k}',\eps'(k')).
\eea
Three measurable quantities have been highlighted - the effect on the probe's frequency 
$\omega$, its wave-vector $\mbf{k}$ and its polarisation $\eps(k)$ and these will be discussed in 
turn.

\subsection{Effects on probe photon polarisation}
\paragraph{Vacuum birefringence} refers to the prediction that the refractive index experienced by a 
probe propagating through regions of intense, but weakly-varying strong fields of amplitude 
$E_{s}$ 
is of the form\cite{Toll:1952rq,baier67a}:
\bea
\tsf{n}^{\parallel,\perp}_{\trm{vac}} = 1 + \frac{(11\mp 3) \alpha}{45\pi} 
\frac{E_{s}^{2}}{\Ecr^{2}}, 
\label{eqn:nvac}
\eea
where the $\parallel$ ($\perp$) indices apply to a probe polarised parallel (perpendicular) to the
strong background. This result may be derived from the Heisenberg-Euler quantum equation of motion,
\be 
   (\partial_{\lambda}\partial^{\lambda} g^{\mu\nu} - \partial^\mu \partial^\nu + \Pi^{\mu\nu}) 
A_\nu = 0 \; .
\ee
A plane wave ansatz for $A_\nu$ implies \emph{two} secular equations or dispersion relations,
\be \label{eqn:MOD.LC}
  k^2 - \Pi_{1,2}(k) = (g^{\mu\nu} - c_{1,2} T^{\mu\nu})k_\mu k_\nu 
  = 0 \; ,
\ee
where $\Pi_{1,2} = c_{1,2} (k,T k)$ are the two nontrivial eigenvalues of the polarisation tensor 
(\ref{eqn:POL.TENSOR}), expressed in terms of the background energy momentum tensor $T^{\mu\nu} = 
F^{\mu}_{\; \alpha} F^{\alpha\nu}$. The dispersion relations (\ref{eqn:MOD.LC}) describe the change 
in light propagation caused by the energy-momentum density stored in the background field and have 
been referred to as modified light-cone conditions\cite{Dittrich:1998fy,Shore:2007um}. They imply 
group velocities different from the vacuum speed of light, $c$, and hence the refractive indices 
(\ref{eqn:nvac}) different from unity, which can be rewritten as 
$\tsf{n}^{\parallel,\perp}_{\trm{vac}} = 1 + \Pi_{1,2}/2 \omega_p^2$, $\omega_p = k^0 c$ being the 
probe frequency.
 
The result for the refractive indices has been shown to hold to all perturbative orders using the 
polarisation operator\cite{Toll:1952rq,Shore:2007um,meuren15} and Heisenberg-Euler Lagrangian 
numerically\cite{ferrando08,boehl15} and analytically\cite{boehl15}. When the pump 
field is space-time dependent as is the case for laser pulses, the effect on the probe is 
calculated by integrating over the inhomogeneous refractive index of the pump 
background\cite{king14a}. There has also been 
recent work indicating finite-time effects in an inhomogeneous background may leave a
detectable signal\cite{hu14b}.

\paragraph{Polarisation flip} is the underlying physical mechanism of vacuum birefringence. The 
term is used when an incoming photon's polarisation vector $\eps^{\mu}$ is ``flipped'' to an 
orthogonal one $\eps^{\prime\,\mu}$ due to real photon-photon scattering. The flip amplitude (for a 
head-on collision of probe and background) after a propagation distance $z$ can be found from the 
Heisenberg-Euler forward scattering amplitude (\ref{eqn:SFI.FWD}) and coincides with the 
birefringence-induced ellipticity \cite{dinu14a},
\be \label{eqn:FLIP.AMP}
  \tsf{e} \equiv \langle \eps' , k| S | \eps , k \rangle = \frac{E_s^2}{\Ecr^{2}} ~\omega_p z ~
  \frac{c_2 - c_1}{2}, \; 
\ee
where $\varepsilon\cdot\varepsilon'=0$. Note the dependence on the \emph{difference} of the low energy constants. This implies that a 
confirmation of vacuum birefringence  would rule out other versions of electrodynamics popular in 
beyond-the-standard-model physics such as Born-Infeld theory, which has $c_1 = 
c_2$\cite{Boillat:1970gw,Plebanski:1970zz,BialynickiBirula:1984tx}. From (\ref{eqn:nvac}), the flip 
amplitude or ellipticity (\ref{eqn:FLIP.AMP}) has the 
equivalent representation
\be \label{eqn:ELLIPTICITY}
  \tsf{e} = \omega_p z ~\frac{\nvac^\perp - \nvac^\parallel}{2} \; ,
\ee
which is proportional to the difference in refractive indices, hence the phase shift between different polarisations.

Detailed calculations have been performed for photons propagating in an arbitrary plane-wave background\cite{baier75a,dinu14a}, and the kinematic low-energy limit relevant for laser-based experiments was found to be consistent with use of the Heisenberg-Euler approach for calculating birefringence and ellipticity\cite{torgrimsson14}. A study of the dependency of the flip and non-flip 
amplitude on spatial and timing jitter and angle of incidence\cite{dinu14b} was performed, with the 
results also being consistent with a previous similar study in the low-energy 
limit\cite{king12}. Both studies\cite{king12,dinu14b} found that 
modelling the background as a focussed paraxial Gaussian beam without taking into account the 
finite pulse duration led to an order of magnitude discrepancy in the number of scattered photons.

\paragraph{Induced ellipticity} is a consequence of birefringence as pointed out in the previous subsection, see (\ref{eqn:FLIP.AMP}) and (\ref{eqn:ELLIPTICITY}). The polarisation of a linearly-polarised probe plane wave can be described with the vector:
\bea
\twovector{\eps^{\parallel}}{\eps^{\perp}} = 
\cos\vphi\twovector{\cos\theta}{\sin\theta}, \label{eqn:ie1}
\eea
where $\vphi$ is the probe phase. If, over some probe phase $\omega_{p}z$ the $\parallel$ 
and $\perp$ 
components experience a different refractive index, then when the phase shift 
$\delta \vphi^{\parallel,\perp} = n^{\parallel,\perp}\omega_{p}z \ll 1$, the polarisation changes 
to:
\bea
\twovector{\eps^{\parallel}}{\eps^{\perp}} = 
  \begin{bmatrix}
    \cos\theta & -\cos\theta~\delta \vphi^{\parallel} \\
    \sin\theta & -\sin\theta~\delta \vphi^{\perp}
  \end{bmatrix}
\twovector{\cos\vphi}{\sin\vphi}, \label{eqn:ie2}
\eea
and the originally linearly-polarised probe is now elliptically polarised. If the background is 
constant, the ellipticity can be written:\cite{dipiazza06}
\bea
\tsf{e} = \omega_{p}z~ \frac{\nvac^{\perp}-\nvac^{\parallel}}{2}\sin 2\theta\; , 
\eea
which generalises (\ref{eqn:ELLIPTICITY}). The induced ellipticity in the interaction of an x-ray probe plane wave of wavelength $\lambda_{p} 
= 0.4\,\trm{nm}$ counterpropagating with a Gaussian pump beam of intensity 
$10^{23}\,\trm{Wcm}^{-2}$ and wavelength $\lambda_{s} = 745\trm{nm}$ focussed to $8\,\mu\trm{m}$ 
was calculated\cite{dipiazza06} to experience an ellipticity of $\tsf{e} \approx 
5\cdot10^{-9}\,\trm{rad}$ when measured at a distance of $0.25\,\trm{m}$ from the pump-probe 
collision. By considering the same pump energy distributed over two 
pump Gaussian laser beams counterpropagating with a Gaussian probe beam, a modest improvement of 
around $\sqrt{2}$ was found, and the near-field induced ellipticity\cite{king10b} 
\bea
\tsf{e} = \frac{2\pi\alpha}{15}\frac{I_{s}}{\Icr}\frac{z_{\trm{eff.}}}{\lambda_{p}}\,\sin2\theta; 
\qquad z_{\trm{eff.}} = \frac{z_{r,p}z_{r,s}}{z_{r,p}+z_{r,s}},
\eea
with the effective interaction length between the two Gaussians $z_{\trm{eff.}}$ depending on the 
probe $z_{r,p}$ and pump $z_{r,s}$ Rayleigh lengths. This agrees with the expressions 
calculated for a monochromatic probe plane wave counterpropagating with a Gaussian 
pump\cite{heinzl06} in the 
limit $z_{r,p}\to \infty$.

\paragraph{Polarisation rotation} is the macroscopic consequence of coherent 
polarisation flipping at the photon level. The effect on the
transverse photon polarisation states in \eqnrefs{eqn:ie1}{eqn:ie2} has the consequence that the 
polarisation angle $\theta$ will rotate as the initially linearly-polarised probe acquires an 
ellipticity. The ellipse traced out by the probe field vector can be seen to be\cite{kingthesis}:
\bea
x^{2} 
-2xy\cos(\delta\vphi^{\perp}-\delta\vphi^{\parallel}) + 
y^{2} = \sin^{2}(\delta\vphi^{\perp}-\delta\vphi^{\parallel}),
\eea
where $x\cos\theta = \eps^{\parallel}$ and $y\sin\theta 
= \eps^{\perp}$. For an x-ray probe counterpropagating with an optical Gaussian pump beam, the 
rotation angle was found to be the same order of magnitude as the induced 
ellipticity\cite{dipiazza06,king10b}.

\subsection{Effects on probe photon wavevector}
On the photon level, four-wave mixing as depicted in \figref{fig:4ph1} can be understood as two 
incoming photons, one from the 
probe and one from the pump, being scattered to two outgoing photons, one being back into the 
pump field and the other being the signal of the vacuum interaction. Conservation of momentum 
permits the scattered photons having a wider transverse distribution than the probe and 
strong 
background, hence allowing one to spatially separate the photon-photon scattering signal from the 
large background of pump and probe laser photons. 

On the classical level, a refractive index
$\nvac$ different from unity, implies altered transmitted wavevectors via Snell's law, and altered transmission $T$ and 
reflection coefficient $R$  via Fresnel's law at perpendicular incidence\cite{jackson75}:
\bea
T = \frac{4\,\nvac}{(1+\nvac)^{2}}; \qquad R = \left(\frac{1-\nvac}{1+\nvac}\right)^{2}.
\eea
If the vacuum refractive index is written as $\nvac = 1 + \delta \nvac$, the effect on 
probe transmission $\sim O(\delta \nvac)$ whereas the effect on reflection $\sim O(\delta 
\nvac^{2})$.
\newline

If the probe beam is considered to be much wider than the pump 
background, the region of polarised vacuum can be considered to ``diffract'' the probe. It is 
well-known that 
the far-field diffracted field is related to the Fourier transform of the aperture 
function\cite{levi68}, and via 
Babinett's principle, this can be related to an integral over the region of  refractive 
index different from unity. We underline the connection of this classical analogue to the intensity form-factor of the 
scattering matrix approach \eqnref{eqn:INTENSITY.FF}.

\paragraph{Vacuum diffraction} was considered in the collision of a plane probe and a 
focussed Gaussian pump beam \cite{dipiazza06}, and extended to to the collision of 
focussed Gaussian probe and pump beams\cite{king10a}. The advantage of this signal is that 
for increasing scattering angle, while the focussed laser 
background is exponentially suppressed, the scattered photon vacuum signal is power-law suppressed. 
In the detector plane then, the number of scattered photons can be calculated in ``measurable'' 
regions, where the signal to noise ratio is much larger than unity. One interesting scenario was 
calculated of colliding two parallel, highly-focussed Gaussian pump beams with a wide 
weakly-focussed Gaussian probe beam, 
such that the photons scattered in the two slit-like polarised regions around the pump beams would 
interfere and hence together form an all-optical double-slit 
experiment\cite{king10a}. For the case of 
two colliding Gaussian pulses, the dependency of the diffracted photon signal on experimental 
parameters such as the total beam power, spatial and timing jitter, angle of collision, pulse 
duration, probe wavelength and focal width has been carried out\cite{king12}. With $10\,\trm{PW}$ 
total laser power split into pump and probe focussed optical pulses, of the order of a few photons 
were predicted to be diffracted into measurable regions on a detector place $1\,\trm{m}$ from the 
interaction centre. 
\begin{figure}
\centering
\includegraphics[width=0.7\linewidth]{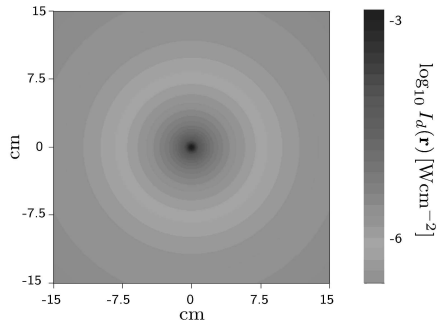}%
\caption{Predicted diffracted electric field in a collision of two counterpropagating Gaussian 
beams. Adapted from \cite{kingthesis}.}
\end{figure}
These results were verified in a study by different authors\cite{tommasini10}, who used a 
different beam model.
The diffraction paradigm was extended from single and double slits to a ``diffraction grating'' of 
having a probe beam diffract off a regular series of pump beams\cite{hatsagortsyan11}. 
Only on positions of the detector where the Bragg condition:
\[
 n q = 2k_{p} \sin\frac{\theta}{2},
\]
for integer $n$, probe wavenumber $k_{p}$, wavenumber of the pump beam structure $q$ and angle 
between incoming and diffracted probe $\theta$, is there constructive interference of the signal of 
scattered photons. Since the addition of diffracted waves occurs at the level of the field, and 
since the number of photons scattered depends upon the total diffracted field squared, there is an 
enhancement in such a set-up proportional to the square of the number of modulation periods. 
Alternatively, rather than using many beams, a single, wide-angle beam diffracting with 
itself at the focus has 
also been studied\cite{monden11}, with the conclusion that the number of diffracted 
photons increases exponentially with the angular aperture. Since only the 
near-field signal was presented, more work is required to determine measurability in this 
scheme.
\newline

The idea of using the diffracted photons' flipped polarisation as well as their altered 
wavevector in an experimental measurement was explored for the wide-angled single-beam 
set-up\cite{monden11}, a single propagating Gaussian beam taking into account higher orders in a 
Hermite-Gauss expansion \cite{paredes14} and has been most recently applied to the upcoming HIBEF 
experiment\cite{karbstein15b}.

\paragraph{Vacuum reflection} refers to the back-scattering of photons in real photon-photon scattering. Static magnetic inhomogeneities of the form of a Lorentzian, 
Gaussian and oscillating Gaussian have been studied\cite{gies13} and more recently static electromagnetic inhomogeneities but most significantly scattering in a Gaussian 
beam\cite{gies15}, although calculations for pulses of a finite duration are still 
to be performed.

\subsection{Effects on probe photon frequency}
\begin{figure}[h!]
\centering
\includegraphics[width=0.35\linewidth]{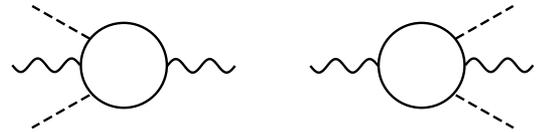}\hspace{0.1\linewidth}
\reflectbox{\includegraphics[width=0.35\linewidth]{figure7}}
\caption{Parametric frequency up-shifting (left) and down-shifting (right) can occur between pump 
and probe through the vacuum interaction.}
\end{figure}
The frequency of probe photons can change via interaction with the polarised vacuum. 
However, this effect is much more difficult to measure experimentally because of the limited range 
of energy and momenta for which it is permitted. Suppose via the four-photon interaction, two 
photons from the strong pump background merge with a probe photon. Then via energy-momentum 
conservation:
\bea
\omega_{p} + \omega_{s,1}+\omega_{s,2} = \omega'; \qquad \mbf{k}_{p} + \mbf{k}_{s,1} + 
\mbf{k}_{s,2} = \mbf{k}', \label{eqn:fwm1}
\eea
but at the same time, the photon must be real to propagate to the detector so $\omega'^{2} = 
\mbf{k}'\cdot\mbf{k}'$. This constrains the allowed frequencies, momenta and angles
that can be combined. 
Similar relations occur for Raman and Brillouin scattering\cite{moloney04}, except all the waves 
here are electromagnetic.

\paragraph{Vacuum parametric frequency-shifting} has been calculated for special beam 
configurations. Combining 
three monochromatic plane waves at right angles, whose wavelengths are $800\,\trm{nm}$, 
$800\,\trm{nm}$ and 
$400\,\trm{nm}$, was predicted to produce a signal that is spatially and 
frequentially (at $267\,\trm{nm}$) separated from the background\cite{lundstroem06}. For respective 
beam powers $0.1\,\trm{PW}$, $0.1\,\trm{PW}$ and $0.5\,\trm{PW}$, taking the interaction region to 
be cuboidal, on average 0.07 photons would be 
frequency-upshifted per collision of the beams, which is predicted to be larger than 
the Compton-scattering background. A signature of the frequency-shifting four-wave mixing 
process 
on the number of total measurable diffracted photons for a collision of two ultra-short 
focussed Gaussian 
pulses was also calculated\cite{king12}. For $10\,\trm{PW}$ total beam power split into a probe 
with wavelength $228\,\trm{nm}$ and duration $2\,\trm{fs}$, as the duration of the 
$910\,\trm{nm}$ pump is reduced to $1\,\trm{fs}$, the total number of diffracted photons is 
predicted to change by around $20\,\%$, equal to one photon per shot. Calculations beyond the 
paraxial approximation
recently performed\cite{fillion-gourdeau15} for two co-propagating beams of different 
frequencies incident on a parabolic mirror suggest $1$-$10\,\trm{PW}$ laser beams are required to 
observe vacuum frequency mixing, although the method of detecting the signal needs to 
be given more attention.

\begin{figure}[h!]
\centering
\includegraphics[width=0.525\linewidth]{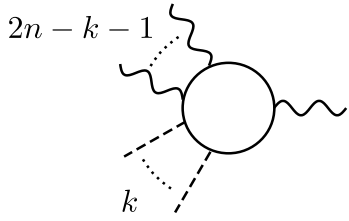}
\caption{Vacuum high-harmonic generation of the $n$th harmonic of the probe via $2n$-photon 
scattering.}\label{fig:vhhg_diag2}
\end{figure}

\begin{figure}[h!]
\centering
\includegraphics[width=0.9\linewidth]{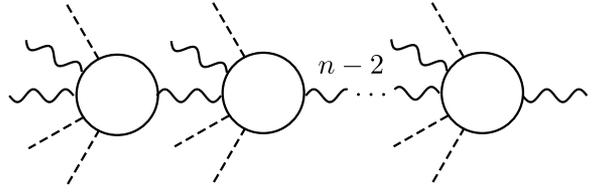}
\caption{Vacuum high-harmonic generation of the $n$th harmonic of the probe via a chain of 
six-photon scattering.}\label{fig:vhhg_shock_diag}
\end{figure}

\paragraph{Vacuum high-harmonic generation} can take place if the colliding laser pulses have the 
same frequency. Then via the four-wave mixing process in \eqnref{eqn:fwm1}, if 
$\omega_{p}=\omega_{s,1}=\omega_{s,2}=\omega$, the signal of the vacuum process has a 
frequency $\omega' = 3\omega$ and so is at the third harmonic of the probe. By considering 
six-, eight- and in general $2n$-wave mixing as depicted in \figref{fig:vhhg_diag2}, it can be seen 
that a harmonic spectrum for the 
vacuum interaction can be produced. 
As each extra interaction between the virtual pair and a laser 
photon is weighted at the amplitude level with a factor $E/\Ecr \ll 1$, higher 
harmonics are in general exponentially suppressed. Nevertheless, the harmonic spectrum produced by 
a standing wave formed of two monochromatic pump laser beams was calculated for 
subcritical 
($E<\Ecr$) strengths where higher harmonic orders $j$ were found\cite{dipiazza05} to follow the 
hierarchy $(E/\Ecr)^{4j}$. In a set-up involving three beams, the minimum power of each laser 
required to scatter one photon was found to be:
\bea
P_{\min} \approx 
33.5\,\frac{\lambda}{1\,\trm{nm}}\frac{w_{0}}{1\,\trm{nm}}\left(\frac{1\trm{fs}}{\tau}\right)^{
1/3}\left(\frac { 1\trm { fs } } { \tau_ {c}}\right)^{2/3}~\trm{GW},
\eea
for typical beam cross-sectional dimension $w_{0}$, interaction duration $\tau$ and coherence time
$\tau_{c}$. The most likely frequency of 
the scattered photon is, however, the fundamental harmonic. The intensity at which a single 
focussed laser pulse will 
begin to produce harmonics via self-interaction has been studied\cite{fedotov07b}, with the 
conclusion that a pulse of $1000\,\trm{nm}$ photons focussed within a cone of angle 
$0.1\,\trm{rad}$ will produce one photon per period at $5\,\cdot \,10^{27}\trm{Wcm}^{-2}$. A recent 
calculation of an alternative route to high-harmonic generation through having many scattering 
events involving low numbers of photons\cite{toll59,bb81,fabrikant82,rozanov93,heyl98,heyl99} (as in 
\figref{fig:vhhg_shock_diag}), 
has recently been suggested to be more efficient. For the collision 
of a Gaussian 
probe at much higher frequency than the background, if the parameter $(64\alpha/105\pi) 
(E_{s}^{3}E_{p}/\Ecr^{4})\omega_{p}\tau_{s}$, where 
$\tau_{s}$ is the duration of the pump, can be made close to unity, 
harmonic generation will dominate, with the
spectrum displaying a power-law behaviour and the appearance of a corresponding electromagnetic 
shock\cite{boehl15}.

\begin{figure}[h!]
\centering
\includegraphics[width=0.55\linewidth]{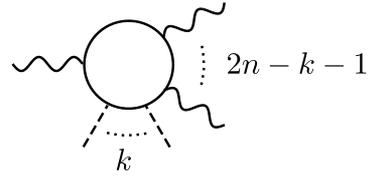}
\caption{An incoming probe photon can split into $k$ outgoing ones, due to interaction with the 
background.} \label{fig:phspl}
\end{figure}

\paragraph{Photon splitting} as depicted in \figref{fig:phspl}, is sometimes thought of as the 
opposite of high-harmonic 
generation, but unlike harmonic generation, the emitted photons can have a continuum of energies.
If one considers splitting to two photons via four-wave mixing then via energy and 
momentum conservation, one possibility is:
\bea
\omega_{p} + \omega_{s} = \omega'_{1} + \omega'_{2}; \qquad \mbf{k}_{p} + \mbf{k}_{s} = 
\mbf{k}'_{1} + \mbf{k}'_{2},
\eea
where now two constraints on these equations are $(\omega'_{1,2})^{2} = 
\mbf{k}'_{1,2}\cdot\mbf{k}'_{1,2}$. The continuum of allowed energies and the possibility for a 
wide angular distribution of emitted photons makes this process worthy of study. The process 
has been comprehensively studied for a probe photon propagating through a plane wave 
background of arbitrary form and polarisation\cite{dipiazza07b}, which was found to depend on 
the two parameters $\eta = \omega_{p}\omega_{s}/m^{2}$ and $\chi=(\omega_{p}/m)(E/\Ecr)$. 
Two events per hour were predicted using $10^{8}$ $250\,\trm{MeV}$ tagged photons per second almost 
counterpropagating with $100\,\trm{fs}$ $10^{15}\,\trm{Wcm}^{-2}$ $1\,\trm{keV}$ XFEL beams 
separated by $93\,\trm{ns}$. Alternatively, two events per hour were also predicted using 
$10^{8}$ $100\,\trm{MeV}$ tagged photons counterpropagating with a 
$1\,\trm{Hz}$ $1\,\trm{eV}$ optical pump of intensity $10^{25}\,\trm{Wcm}^{-2}$. The conclusion was 
that a different experimental set-up must be considered if this effect is to be observed in the 
near future\cite{dipiazza07b}.

\subsection{Effects on probe pulse form}
In addition to the effects on single photons, one can consider the consequence of real 
photon-photon scattering on the propagation of an ensemble of photons. A probe laser pulse can be 
understood as a superposition of photons with a range of frequencies and phases. From the study of 
nonlinear dispersive media, it is well known that a refractive index that depends on a probe's 
intensity directly or indirectly can lead to pulse shape effects\cite{moloney04}. In 
particular for the interaction with vacuum, probe pulse effects can occur if the 
next-to-leading order effect of a probe-dependent refractive index is taken into account. 

\paragraph{Nonlinear phase shift} is a term used to denote the relative difference in phases 
between parts of a probe beam that have experienced different vacuum refractive indices. For 
a constant refractive index, the relative phase difference compared to a unitary refractive 
index is:
\[
 \delta \phi = (\nvac - 1)\omega_{p}z,
\]
where $\omega_{p} z$ is the phase over which $\delta\phi$ has been accrued. For two 
counterpropagating initially monochromatic plane waves, with the envisaged ELI parameters of 
$800\,\trm{nm}$ wavelength, 
$10^{25}\,\trm{Wcm}^{-2}$ intensity, $10\,\trm{fs}$ duration and $10\,\mu\trm{m}$ focal spot 
diameter, 
a phase shift of the order of $\delta\phi\approx 10^{-7}\,\trm{rad}$ has been 
calculated\cite{ferrando08,tommasini08}. This nonlinear phase shift can be enhanced by using 
multiple crossings of the interacting beams. For $N_{r}$ reflections from plasma mirrors of 
reflectivity $R_{\trm{mir}}$ of two beams crossing each other at an angle $\theta_{c}$, the gain 
factor has been calculated to be\cite{tommasini09}:
\[
 \sin^{4}\left(\frac{\theta_{c}}{2}\right)~\sum_{n=0}^{N_{r}+1}R^{n}_{\trm{mir}}.
\]
The measurement of this phase shift using Fourier imaging has also been explored\cite{tajima11a}.

\paragraph{Vacuum self-focusing} is an analogue to the well-known plasma self-focussing or 
``Benjamin-Weir'' instability\cite{moloney04} in which there is positive feedback between a 
refractive index increasing the intensity of a pulse via focussing, and a higher 
intensity resulting from that focussing in turn increasing the refractive index. Mutual channelling 
of counterpropagating laser pulses and 
large-scale focussing have been considered, but either $\trm{YW}$ powers are predicted as 
necessary\cite{rozanov93} or intensities above critical\cite{fedotov07}, before which vacuum 
pair-creation would have set in. In considering the idealised geometry of a Gaussian plane wave 
probe pulse counterpropagating and interacting via six-wave mixing with a much slower varying pump, 
the probe-dependent refractive index:
\bea
\nvac^{\parallel} = 1 + \frac{\alpha}{\pi}\frac{E_{s}^{2}}{\Ecr^{2}}\left[\frac{8}{45} + 
\frac{64}{105}\frac{E_{s}}{\Ecr}\frac{E_{p}}{\Ecr}\right]
\eea
was predicted to lead to the generation of a shock wave, a signature of self-focussing, when the 
phase difference due to 
the probe-dependent refractive index tended to a quarter-wavelength\cite{boehl15}.

\begin{figure}[h]
\begin{center}
\includegraphics[width=0.95\linewidth]{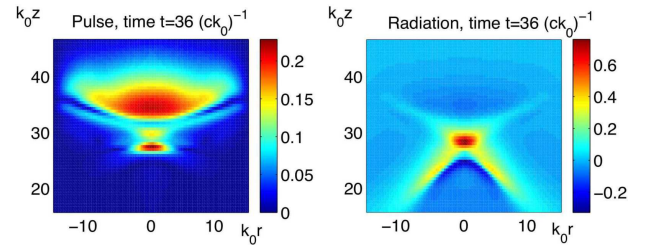}
\caption{\label{fig:cerenkov} Cerenkov-like radiation (right) generated by pulse collapse 
into photon bullets (left) against longitudinal $z$ and transverse $r$ co-ordinates of an 
initially Gaussian pulse of central wavenumber $k_{0}$. Reproduced with 
permission\cite{marklund03}.}
\end{center}
\end{figure}

\paragraph{Pulse collapse} is predicted to occur for high-intensity probe pulses propagating 
through an even higher-intensity background. The wave-equation for the probe can be recast as a 
nonlinear Schr\"odinger 
equation\cite{moloney04} with the consequence that the pulse envelope becomes spacetime dependent, 
even if assumed initially homogeneous. Unlike typical optically-nonlinear dispersive media, the 
nonlinearity of the vacuum is ``formed'' by the pump laser background, which is then probed by a 
second pulse. 
Even when the leading-order effect on the probe is a nonlinear refractive index that is 
independent of the probe pulse, because of its effect on the pump's evolution, it can 
indirectly effect the probe's propagation. This interplay between a Gaussian probe 
distribution propagating 
through a radiation gas has been demonstrated to lead to self-focussing and collapse of the 
probe into ``photon bullets'', thereby driving acoustic waves\cite{marklund03} as demonstrated in 
\figref{fig:cerenkov}. 
Depending on 
initial parameters, probe collapse can occur before or after the critical Schwinger limit is 
reached\cite{marklund04}. 

\subsection{Finite time effects}
Similar to the case for regular plasmas, there are effects on the probe when propagating through 
regions of the polarised ``vacuum plasma'' that do not persist long enough to be directly detected.

\paragraph{Photon acceleration} is well-known from plasma physics\cite{wilks89} and corresponds 
to the frequency downshift (upshift) as probe photons traverse an increasing (decreasing) plasma 
gradient. The possibility of measuring this effect in vacuum has been considered for a probe photon 
propagating almost parallel with a pump pulse\cite{mendonca06}, with a frequency up (down) shift 
occurring at the rear (front) of the pump beam.

\paragraph{High harmonic generation} can also occur due to the inhomogeneity of the pump pulse 
background, in an effect distinct from standard vacuum high harmonic generation. For 
a probe pulse counterpropagating with a slowly-varying background, this is predicted to 
occur at finite time during overlap of the 
probe and pump pulses at an order earlier (via four-photon scattering), than for those photons that 
reach a detector (via six-photon scattering)\cite{boehl14}. This finite-time signal disappears when 
the probe and pump pulses are well-separated again, but is calculated to dominate the signal of 
frequency-shifted photons when the pulses overlap in this set-up if
$(E_{s}/\Ecr)^{2}\omega_{p}\tau_{s}\ll1$ for strong-pulse duration $\tau_{s}$.

\paragraph{Gradient-dependent vacuum refractive index} is a way to describe the addition to the 
standard predicted vacuum refractive index that occurs when the pump laser is time-varying. This 
has been calculated for a probe propagating through the electric/magnetic antinode of a 
pump standing wave\cite{hu14b}. The change in vacuum refractive index $\Delta\nvac$ can be 
written in the form:
\bea
\Delta\nvac^{\parallel,\perp}(\vphi) = \frac{E_{p}}{E_{p}'}
\nvac^{\parallel,\perp \, \prime}(\vphi).
\eea
In a set-up of two colliding plane waves with no transverse structure, it was shown that this term 
is a surface term and is zero initially and finally, when the probe and background are 
well-separated\cite{boehl14}. The contact term was also noted in a recent study of polarisation 
flipping in arbitrary plane waves\cite{dinu14a}. Although it has been suggested this part of the 
interaction could be a useful probe of dark matter particles\cite{hu14b}, a consistent 
finite-time calculation has yet to be performed to establish the nature of this effect.

\subsection{Non-perfect vacua} 
In any realistic experiment, the vacuum will be synthetic and hence imperfect. Residue particles in 
interaction chambers will also be affected by intense laser pulses and can produce a source of 
background that may obscure the measurement of real photon-photon scattering. The Cotton-Mouton 
effect, in which a dilute gas becomes birefringent in the presence of an electromagnetic wave is 
just one such example\cite{pvlas12}. In light of this, various proposals have been considered that 
instead
use an altered vacuum to enhance the signal of vacuum polarisation.
\newline

\paragraph{Resonant Cavities} can be employed in order to increase the sensitivity of 
whatever eigenfrequencies are resonant for that particular cavity\cite{jackson75}. For example, a 
cavity can be designed such that the frequency that is generated by vacuum four-wave mixing of two 
modes of the cavity, is resonant. This idea has been studied for the $\trm{TE}_{01}$ modes of such 
a cavity and the growth of the mixing signal in the form of the longitudinal standing-wave magnetic 
field, found 
to increase linearly with time\cite{brodin01} as
\[
 B_{3}(t) = \frac{itV}{2\omega_{3}}\,B_{1}^{2}B_{2}^{\ast},
\]
for source magnetic standing wave strengths $B_{1}, B_{2}$ and coupling constant $V$. The vacuum 
signal was predicted to be detectable if 
an electric field $2\times10^{-8}$ times the critical Schwinger field was employed with a 
superconducting cavity with a resistance of $1\,\trm{n}\Omega$ and a resonant, 
vacuum-mixing frequency of $13.2\,\mu\trm{eV}$. This idea was refined\cite{eriksson04} and the 
prediction made that $18$ photons can be produced by a magnetic field of around $0.28\,\trm{T}$ in 
a 
cylindrical cavity of length $2.5\,\trm{m}$, radius $25\,\trm{cm}$ and quality factor 
$4\times10^{10}$.
\newline

\paragraph{Real plasmas} already have a 
refractive index different from unity, and this can combine with the shift 
of the refractive index due to vacuum polarisation and 
lead to an enhancement. The system of equations by 
Akhiezer and Polovin\cite{akhiezer56} for the propagation of a 
circularly-polarised plane wave through a cold 
collisionless plasma was updated to include the vacuum 
current in Maxwell's equations and also take into 
account collisions\cite{dipiazza07}. For the collisionless case, the 
modified refractive index of the combined system was found to be:
\[
 \tsf{n} = \sqrt{\tsf{n}_{\trm{pl}}^{2} + 
\frac{1}{4}\delta\tsf{n}_{\trm{vac}}^{\perp}(1-\tsf{n}_{\trm{pl}}^{2})^{2}}
\]
with $\tsf{n}_{\trm{pl}}$ the plasma refractive index and $\delta \nvac^{\perp} = \nvac^{\perp}-1$ 
as 
defined in \eqnref{eqn:nvac}. Another detectable signal of photon-photon scattering has been 
calculated to exist when an overdense plasma channel is subjected to an intense laser 
beam\cite{shen03}. In addition, the altered dispersion relation for electromagnetic waves due to 
vacuum polarisation effects in a strongly-magnetised cold plasma has 
been calculated\cite{brodin06,lundin07,shukla08}, which is particularly relevant for the dynamics 
of strongly-magnetised neutron stars.
%
%
%
%
%
%
\section{Summary}
There has been a proliferation of labels to describe polarisation effects of the quantum vacuum due 
to intense laser pulses. However, as we have discussed, all of 
these are manifestations of the QED prediction that real photons can scatter off one 
another. The commonality of the main approaches of describing real photon-photon 
scattering, through calculation of the polarisation operator, scattering matrix elements and 
Heisenberg-Euler-modified Maxwell equations, has been made manifest. Many signals of this 
long-predicted phenomenon, whether at the level of individual photons or at the level of 
electromagnetic fields, have been calculated and found measurable in experiments using 
high-intensity laser pulses. This implies that the first measurement of real 
photon-photon scattering will finally be performed in the near future.

\bibliographystyle{JHEP}
\bibliography{current2}

\providecommand{\noopsort}[1]{}
\providecommand{\href}[2]{#2}\begingroup\raggedright\begin{thebibliography}{10}

\bibitem{danson15}
C.~Danson, D.~Hillier, N.~Hopps, and D.~Neely, {\it Petawatt class lasers
  worldwide},  {\em High Power Laser Science and Engineering} {\bf 3} (2015).

\bibitem{marklund_review06}
M.~Marklund and P.~K. Shukla, {\it Nonlinear collective effects in
  photon-photon and photon-plasma interactions},  {\em Rev. Mod. Phys.} {\bf
  78} (2006) 591.

\bibitem{dipiazza12}
A.~{Di Piazza} et~al., {\it Extremely high-intensity laser interactions with
  fundamental quantum systems},  {\em Rev. Mod. Phys.} {\bf 84} (2012)
  1177--1228.

\bibitem{bernard00}
D.~Bernard et~al., {\it Search for stimulated photon-photon scattering in
  vacuum},  {\em Eur. Phys. J. D} {\bf 10} (2000) 141--145.

\bibitem{rizzo13}
A.~Cad\`ene, P.~Berceau, M.~Fouch\'e, R.~Battesti, and C.~Rizzo, {\it {Vacuum
  magnetic linear birefringence using pulsed fields: status of the BMV
  experiment}},  {\em Eur. Phys. J.} {\bf D68} (2014) 16,
  [\href{http://xxx.lanl.gov/abs/1302.5389}{{\tt arXiv:1302.5389}}].

\bibitem{DellaValle:2014xoa}
F.~Della~Valle, E.~Milotti, A.~Ejlli, G.~Messineo, L.~Piemontese, G.~Zavattini,
  U.~Gastaldi, R.~Pengo, and G.~Ruoso, {\it {First results from the new PVLAS
  apparatus: A new limit on vacuum magnetic birefringence}},  {\em Phys. Rev.}
  {\bf D90} (2014) 092003.

\bibitem{schlenvoigt15}
H.~Schlenvoigt, T.~Heinzl, U.~Schramm, T.~Cowan, and R.~Sauerbrey, ``Detecting
  vacuum birefringence with {X}-ray free electron lasers and high-power optical
  lasers: A feasibility study.'' submitted to Phys.\ Scripta, 2015.

\bibitem{Weinberg:1995mt}
S.~Weinberg, {\em {The Quantum theory of fields. Vol. 1: Foundations}}.
\newblock Cambridge University Press, 2005.

\bibitem{peskin95}
M.~E. Peskin and D.~V. Schroeder, {\em An Introduction to Quantum Field
  Theory}.
\newblock Westview, 1995.

\bibitem{gusynin99}
V.~P. Gusynin and I.~A. Shovkovy, {\it Derivative expansion of the effective
  action for {QED} in 2+1 and 3+1 dimensions},  {\em J. Math. Phys.} {\bf 40}
  (1999) 5406.

\bibitem{dunne99}
G.~V. Dunne and T.~M. Hall, {\it Borel summation of the derivative expansion
  and effective actions},  {\em Phys. Rev. D} {\bf 60} (1999) 065002.

\bibitem{heisenberg36}
W.~Heisenberg and H.~Euler, {\it Folgerungen aus der {D}iracschen {T}heorie des
  {P}ositrons},  {\em Z. Phys.} {\bf 98} (1936) 714.

\bibitem{weisskopf36}
V.~Weisskopf, {\it {{\"U}ber die Elektrodynamik des Vakuums auf Grund der
  Quantentheorie des Elektrons}},  {\em Kgl. Danske Videnskab. Selskab,
  Mat.-fys. Medd.} {\bf 14} (1936) 6.

\bibitem{schwinger51}
J.~Schwinger, {\it On gauge invariance and vacuum polarization},  {\em Phys.
  Rev.} {\bf 82} (1951) 664--679.

\bibitem{Sauter:1931zz}
F.~Sauter, {\it {{\"U}ber das {V}erhalten eines {E}lektrons im homogenen
  elektrischen {F}eld nach der relativistischen {T}heorie {D}iracs}},  {\em Z.
  Phys.} {\bf 69} (1931) 742--764.

\bibitem{landau4}
V.~B. Berestetskii, E.~M. Lifshitz, and L.~P. Pitaevskii, {\em Quantum
  Electrodynamics (second edition)}.
\newblock Butterworth-Heinemann, Oxford, 1982.

\bibitem{Toll:1952rq}
J.~S. Toll, {\em {The dispersion relation for light and its application to
  problems involving electron pairs}}.
\newblock PhD thesis, Princeton U., 1952.

\bibitem{narozhny69}
N.~B. Narozhny\u{\i}, {\it Propagation of plane electromagnetic waves in a
  constant field},  {\em Sov. Phys. JETP} {\bf 28} (1969) 371--374.

\bibitem{baier75a}
V.~N. Ba\u{\i}er, A.~I. Mil'shte\u{\i}n, and V.~M. Strakhovenko, {\it
  Interaction between a photon and an intense electromagnetic wave},  {\em Sov.
  Phys. JETP} {\bf 42} (1976), no.~6 961--965.

\bibitem{meuren13}
S.~Meuren, C.~H. Keitel, and A.~Di~Piazza, {\it {Polarization operator for
  plane-wave background fields}},  {\em Phys. Rev.} {\bf D88} (2013) 013007.

\bibitem{Dittrich:2000zu}
W.~Dittrich and H.~Gies, {\it {Probing the quantum vacuum. Perturbative
  effective action approach in quantum electrodynamics and its application}},
  {\em Springer Tracts Mod. Phys.} {\bf 166} (2000) 1--241.

\bibitem{Akhiezer:1965}
A.~Akhiezer and V.~Berestetskii, {\em Quantum Electrodynamics}.
\newblock Interscience, New York, 1965.

\bibitem{dipiazza13}
A.~Di~Piazza, {\it On refractive processes in strong laser field quantum
  electrodynamics},  {\em Ann. Phys.} {\bf 338} (2013) 302.

\bibitem{delphenich06}
D.~H. Delphenich, {\it Nonlinear optical analogies in quantum electrodynamics},
   \href{http://xxx.lanl.gov/abs/hep-ph/0610088}{{\tt hep-ph/0610088}}.

\bibitem{narozhny15}
N.~B. Narozhny and A.~M. Fedotov, {\it Extreme light physics},  {\em
  Contemporary Physics} {\bf 56} (2015) 249--268.

\bibitem{baier67a}
R.~Baier and P.~Breitenlohner {\em Acta Phys. Austriaca} {\bf 25} (1967) 212.

\bibitem{Dittrich:1998fy}
W.~Dittrich and H.~Gies, {\it {Light propagation in nontrivial QED vacua}},
  {\em Phys. Rev.} {\bf D58} (1998) 025004,
  [\href{http://xxx.lanl.gov/abs/hep-ph/9804375}{{\tt hep-ph/9804375}}].

\bibitem{Shore:2007um}
G.~M. Shore, {\it {Superluminality and UV completion}},  {\em Nucl. Phys.} {\bf
  B778} (2007) 219--258, [\href{http://xxx.lanl.gov/abs/hep-th/0701185}{{\tt
  hep-th/0701185}}].

\bibitem{meuren15}
S.~Meuren, K.~Z. Hatsagortsyan, C.~H. Keitel, and A.~Di~Piazza, {\it
  Polarization-operator approach to pair creation in short laser pulses},  {\em
  Phys. Rev. D} {\bf 91} (2015) 013009.

\bibitem{ferrando08}
A.~Ferrando, H.~Michinel, M.~Seco, and D.~Tommasini, {\it Nonlinear phase shift
  from photon-photon scattering in vacuum},  {\em Phys. Rev. Lett.} {\bf 99}
  (2007) 150404.

\bibitem{boehl15}
P.~B\"ohl, B.~King, and H.~Ruhl, {\it Vacuum high-harmonic generation in the
  shock regime},  {\em Phys. Rev. A} {\bf 92} (2015) 032115.

\bibitem{king14a}
B.~King, P.~B\"ohl, and H.~Ruhl {\em Phys. Rev. D} {\bf 90} (2014) 065018.

\bibitem{hu14b}
H.~Hu and J.~Huang, {\it Modified light-cone condition via vacuum polarization
  in a time-dependent field},  {\em Phys. Rev. A} {\bf 90} (2014) 062111.

\bibitem{dinu14a}
V.~Dinu, T.~Heinzl, A.~Ilderton, M.~Marklund, and G.~Torgrimsson, {\it Vacuum
  refractive indices and helicity flip in strong-field {QED}},  {\em Phys. Rev.
  D} {\bf 89} (2014) 125003.

\bibitem{Boillat:1970gw}
G.~Boillat, {\it {Nonlinear electrodynamics - {Lagrangians} and equations of
  motion}},  {\em J. Math. Phys.} {\bf 11} (1970) 941--951.

\bibitem{Plebanski:1970zz}
J.~Plebanski, {\it {Lectures on Nonlinear Electrodynamics}}, . {Lectures on
  Nonlinear Electrodynamics}, Nordita report RX-476.

\bibitem{BialynickiBirula:1984tx}
I.~Bia{\l}ynicki-Birula, {\it {Nonlinear {E}lectrodynamics: {V}ariations on a
  {T}heme by {B}orn and {I}nfeld}}, . in: \textit{Quantum Theory of Particles
  and Fields}, B.~Jancewicz and J.~Lukierski, eds., World Scientific,
  Singapore, 1983.

\bibitem{torgrimsson14}
G.~Torgrimsson, {\it {Ellipticity induced in vacuum birefringence}},  in {\em
  {Theory and Experiment for Hadrons on the Light-Front (Light Cone 2014)
  Raleigh, North Carolina, USA, May 26-30, 2014}}.
\newblock \href{http://xxx.lanl.gov/abs/1409.8069}{{\tt arXiv:1409.8069}}.

\bibitem{dinu14b}
V.~Dinu, T.~Heinzl, A.~Ilderton, M.~Marklund, and G.~Torgrimsson, {\it Photon
  polarization in light-by-light scattering: {Finite size effects}},  {\em
  Phys. Rev. D} {\bf 90} (2014) 045025.

\bibitem{king12}
B.~King and C.~H. Keitel, {\it Photon-photon scattering in collisions of laser
  pulses},  {\em New J. Phys.} {\bf 14} (2012) 103002.

\bibitem{dipiazza06}
A.~{Di Piazza}, K.~Z. Hatsagortsyan, and C.~H. Keitel, {\it Light diffraction
  by a strong standing electromagnetic wave},  {\em Phys. Rev. Lett.} {\bf 97}
  (2006) 083603.

\bibitem{king10b}
B.~King, A.~{Di Piazza}, and C.~H. Keitel, {\it Double-slit vacuum polarisation
  effects in ultra-intense laser fields},  {\em Phys. Rev. A} {\bf 82} (2010)
  032114.

\bibitem{heinzl06}
T.~Heinzl et~al., {\it On the observation of vacuum birefringence},  {\em Opt.
  Commun.} {\bf 267} (2006) 318--321.

\bibitem{kingthesis}
B.~King, {\em Vacuum Polarisation Effects in Intense Laser Fields}.
\newblock PhD thesis, University of Heidelberg
  \url{http://www.ub.uni-heidelberg.de/archiv/10846}, 2010.

\bibitem{jackson75}
J.~D. Jackson, {\em Classical Electrodynamics}.
\newblock John Wiley \& Sons, Inc., New York, 1975.

\bibitem{levi68}
L.~Levi, {\em Applied Optics}.
\newblock John Wiley \& Sons, Inc., New York, 1968.

\bibitem{king10a}
B.~King, A.~D. Piazza, and C.~H. Keitel, {\it A matterless double-slit},  {\em
  Nature Photon.} {\bf 4} (2010) 92--94.

\bibitem{tommasini10}
D.~Tommasini and H.~Michinel, {\it Light by light diffraction in vacuum},  {\em
  Phys. Rev. A (R)} {\bf 82} (2010) 011803.

\bibitem{hatsagortsyan11}
G.~Y. Kryuchkyan and K.~Z. Hatsagortsyan, {\it Bragg scattering of light in
  vacuum structured by strong periodic fields},  {\em Phys. Rev. Lett.} {\bf
  107} (2011) 053604.

\bibitem{monden11}
Y.~Monden and R.~Kodama, {\it Enhancement of laser interaction with vacuum for
  a large angular aperture},  {\em Phys. Rev. Lett.} {\bf 107} (2011) 073602.

\bibitem{paredes14}
A.~Paredes, D.~Novoa, and D.~Tommasini, {\it Self-induced mode mixing of
  ultraintense lasers in vacuum},  {\em Phys. Rev. A} {\bf 90} (2014) 063803.

\bibitem{karbstein15b}
F.~Karbstein, H.~Gies, M.~Reuter, and M.~Zepf, {\it {Vacuum birefringence in
  strong inhomogeneous electromagnetic fields}},
  \href{http://xxx.lanl.gov/abs/1507.0108}{{\tt arXiv:1507.0108}}.

\bibitem{gies13}
H.~Gies, F.~Karbstein, and N.~Seegert {\em New J. Phys.} {\bf 15} (2013)
  083002.

\bibitem{gies15}
H.~Gies, F.~Karbstein, and N.~Seegert, {\it Quantum reflection of photons off
  spatio-temporal electromagnetic field inhomogeneities},  {\em New Journal of
  Physics} {\bf 17} (2015) 043060.

\bibitem{moloney04}
J.~V. Moloney and A.~C. Newell, {\em Nonlinear Optics}.
\newblock Westview Press, Oxford, UK, 2004.

\bibitem{lundstroem06}
E.~Lundstr{\"o}m et~al., {\it Using high-power lasers for detection of elastic
  photon-photon scattering},  {\em Phys. Rev. Lett.} {\bf 96} (2006) 083602.

\bibitem{fillion-gourdeau15}
F.~Fillion-Gourdeau, C.~Lefebvre, and S.~MacLean, {\it Scheme for the detection
  of mixing processes in vacuum},  {\em Phys. Rev. A} {\bf 91} (2015) 031801.

\bibitem{dipiazza05}
A.~{Di Piazza}, K.~Z. Hatsagortsyan, and C.~H. Keitel, {\it Harmonic generation
  from laser-driven vacuum},  {\em Phys. Rev. D} {\bf 72} (2005) 085005.

\bibitem{fedotov07b}
A.~M. Fedotov and N.~B. Narozhny, {\it Generation of harmonics by a focused
  laser beam in the vacuum},  {\em Phys. Lett. A} {\bf 362} (2007) 1--5.

\bibitem{toll59}
M.~Lutzky and J.~S. Toll, {\it Formation of discontinuities in classical
  nonlinear electrodynamics},  {\em Phys. Rep.} {\bf 113} (1959) 1649.

\bibitem{bb81}
Z.~Bialynicka-Birula, {\it Nonlinear phenomena in the propagation of
  electromagnetic waves in the magnetized vacuum},  {\em Physica D} {\bf 2}
  (1981) 513.

\bibitem{fabrikant82}
V.~V. Zheleznyakov and A.~L. Fabrikant, {\it Electromagnetic shock waves in a
  magnetized vacuum},  {\em Sov. Phys. JETP} {\bf 55} (1982) 794.

\bibitem{rozanov93}
N.~N. Rozanov, {\it Four-wave interactions of intense radiation in vacuum},
  {\em Sov. Phys. JETP} {\bf 76} (1993) 991.

\bibitem{heyl98}
J.~S. Heyl and L.~Hernquist, {\it Electromagnetic shocks in strong magnetic
  fields},  {\em Phys. Rev. D} {\bf 58} (1998) 043005.

\bibitem{heyl99}
J.~S. Heyl and L.~Hernquist, {\it Nonlinear {QED} effects in strong-field
  magnetohydrodynamics},  {\em Phys. Rev. D} {\bf 59} (1999) 045005.

\bibitem{dipiazza07b}
A.~{Di Piazza}, A.~I. Milstein, and C.~H. Keitel, {\it Photon splitting in a
  laser field},  {\em Phys. Rev. A} {\bf 76} (2007) 032103.

\bibitem{tommasini08}
D.~Tommasini et~al., {\it Detecting photon-photon scattering in vacuum at
  exawatt lasers},  {\em Phys. Rev. A} {\bf 77} (2008) 042101.

\bibitem{tommasini09}
D.~Tommasini, A.~Ferrando, M.~Humberto, and M.~Seco, {\it Precision tests of
  {QED} and non-standard models by searching photon-photon scattering in vacuum
  with high power lasers},  {\em J. High Energy Phys.} {\bf 11} (2009) 43.

\bibitem{tajima11a}
K.~Homma, D.~Habs, and T.~Tajima, {\it Probing vacuum birefringence by
  phase-contrast fourier imaging under fields of high-intensity lasers},  {\em
  Appl. Phys. B-Lasers O.} {\bf 104} (2011) 769--782.

\bibitem{fedotov07}
A.~M. Fedotov {\em Proc. SPIE} {\bf 6726} (2007) 67261D.

\bibitem{marklund03}
M.~Marklund, G.~Brodin, and L.~Stenflo, {\it {Electromagnetic Wave Collapse in
  a Radiation Background}},  {\em Phys. Rev. Lett.} {\bf 91} (2003) 163601.

\bibitem{marklund04}
M.~Marklund, B.~Eliasson, and P.~K. Shukla {\em Sov. Phys. JETP} {\bf 79}
  (2004) 208.

\bibitem{wilks89}
S.~C. Wilks, J.~M. Dawson, W.~B. Mori, T.~Katsouleas, and M.~E. Jones, {\it
  Photon accelerator},  {\em Phys. Rev. Lett.} {\bf 62} (1989) 2600--2603.

\bibitem{mendonca06}
J.~T. Mendonca et~al., {\it Photon acceleration in vacuum},  {\em Phys. Lett.
  A} {\bf 359} (2006) 700--704.

\bibitem{boehl14}
B.~King, P.~B\"ohl, and H.~Ruhl {\em Phys. Rev. D} {\bf 90} (2014) 065018.

\bibitem{pvlas12}
G.~Zavattini et~al., {\it Measuring the magnetic birefringence of vacuum: {The
  PVLAS} experiment},  {\em Int. J. Mod. Phys. A} {\bf 27} (2012) 1260017.

\bibitem{brodin01}
G.~Brodin, M.~Marklund, and L.~Stenflo, {\it {Proposal for Detection of QED
  Vacuum Nonlinearities in Maxwell’s Equations by the Use of Waveguides}},
  {\em Phys. Rev. Lett.} {\bf 87} (2001) 171801.

\bibitem{eriksson04}
D.~Eriksson, G.~Brodin, M.~Marklund, and L.~Stenflo, {\it Possibility to
  measure elastic photon-photon scattering in vacuum},  {\em Phys. Rev. A} {\bf
  70} (2004) 013808.

\bibitem{akhiezer56}
A.~I. Akhiezer and R.~V. Polovin {\em JETP} {\bf 3} (1956) 696.

\bibitem{dipiazza07}
A.~{Di Piazza}, K.~Z. Hatsagortsyan, and C.~H. Keitel, {\it Enhancement of
  vacuum-polarization effects in a plasma},  {\em Phys. Plasmas} {\bf 14}
  (2007) 032102.

\bibitem{shen03}
B.~Shen, M.~Y. Yu, and X.~Wang, {\it {Photon–photon scattering in a plasma
  channel}},  {\em Physics of Plasmas} {\bf 10} (2003) 4570--4571.

\bibitem{brodin06}
G.~Brodin, M.~Marklund, L.~Stenflo, and P.~K. Shukla, {\it Dispersion relation
  for electromagnetic wave propagation in a strongly magnetized plasma},  {\em
  New Journal of Physics} {\bf 8} (2006) 16.

\bibitem{lundin07}
J.~Lundin, L.~Stenflo, G.~Brodin, M.~Marklund, and P.~K. Shukla, {\it
  Circularly polarized waves in a plasma with vacuum polarization effects},
  {\em Physics of Plasmas} {\bf 14} (2007) 6.

\bibitem{shukla08}
P.~K. Shukla and L.~Stenflo, {\it Dispersion relations for electromagnetic
  waves in a dense magnetized plasma},  {\em Journal of Plasma Physics} {\bf
  74} (2008) 719--723.

\end{thebibliography}\endgroup
\end{document}